# Arrays of Ag split-ring resonators coupled to InGaAs single-quantum-well gain

Nina Meinzer,[1,2,*] Matthias Ruther,[1,2] Stefan Linden,[1,2] Costas M. Soukoulis,[3] Galina Khitrova,[4] Joshua Hendrickson,[4] Joshua D. Olitsky,[4] Hyatt M. Gibbs,[4] and Martin Wegener[1,2]

[1] *Institut für Nanotechnologie, Karlsruhe Institute of Technology (KIT), Postfach 3640, D-76021 Karlsruhe, Germany*
[2] *Institut für Angewandte Physik and DFG-Center for Functional Nanostructures (CFN), Karlsruhe Institute of Technology (KIT), D-76128 Karlsruhe, Germany*
[3] *Ames Laboratory and Department of Physics and Astronomy, Iowa State University, Ames, Iowa 50011, U.S.A. and Research Center of Crete, and Department of Materials Science and Technology, 71110 Heraklion, Crete, Greece*
[4] *College of Optical Sciences, The University of Arizona, Tucson, AZ 85721, U.S.A.*
*\*Nina.Meinzer@kit.edu*

**Abstract:** We study arrays of silver split-ring resonators operating at around 1.5-µm wavelength coupled to an MBE-grown single 12.7-nm thin InGaAs quantum well separated only 4.8 nm from the wafer surface. The samples are held at liquid-helium temperature and are pumped by intense femtosecond optical pulses at 0.81-µm center wavelength in a pump-probe geometry. We observe much larger relative transmittance changes (up to about 8%) on the split-ring-resonator arrays as compared to the bare quantum well (not more than 1-2%). We also observe a much more rapid temporal decay component of the differential transmittance signal of 15 ps for the case of split-ring resonators coupled to the quantum well compared to the case of the bare quantum well, where we find about 0.7 ns. The latter observation is ascribed to the Purcell effect that arises from the evanescent coupling of the split-ring resonators to the quantum-well gain. All experimental results are compared with a recently introduced analytical toy model that accounts for this evanescent coupling, leading to excellent overall qualitative agreement.

## 1. Introduction

At near-infrared and visible frequencies, losses of metal-based optical metamaterials are very large due to the fact that the intrinsic free-electron metal losses increase drastically when even remotely approaching the metal plasma frequency (see, *e.g.*, the reviews [1-3]). For many of the envisioned applications of metamaterials such as in perfect lenses [4] or in transformation optics [5], low-loss or even zero-loss metamaterials are required. Theoretical calculations have shown that introducing optical gain is a possible remedy [6-15].

So far, to the best of our knowledge, only two experiments [16,17] along these lines have been published. Ref. 16 has used an approximately 1-µm thick film of 3.2-nm diameter PbS quantum dots spun onto an array of complementary split-ring resonators (with two slits) operating at around 1.0-µm wavelength. At room temperature and upon continuous-wave optical pumping, they observe induced transmittance changes on the order of 1% [see their [16] Fig. 2(d)]. Ref. 17 has employed Rh800 dye molecules embedded in an epoxy filled into a double-fishnet-type negative-index metamaterial operating at around 0.7-µm wavelength. They measure an increased relative transmittance on the order of 100% [see their [17] Fig. 3(b)] upon pulsed optical pumping at room temperature and infer zero loss at some wavelength by comparison with detailed numerical calculations [17].

In the present work, we focus on optically-pumped epitaxially-grown single 12.7-nm thin InGaAs semiconductor quantum wells in close proximity to a layer of 30-nm thin silver split-ring resonators operating at around 1.5-µm wavelength. To maximize the optical gain, we cool the samples to helium temperature. We use semiconductor quantum wells because they open perspectives towards electrical carrier injection (as in any semiconductor laser) and because they are long-term photo-stable – in sharp contrast to, *e.g.*, dye molecules.

Before turning to our experiments, we would like to provide an intuitive qualitative discussion of what effects can be expected. In this discussion, the coupling strength between a metamaterial layer and the gain is a determining factor. We start our reasoning with the case of no coupling and continue towards weak and strong coupling.

To illustrate the limit of no coupling, let us consider a metamaterial layer and a gain medium separated by the absurd distance of one meter. It is clear that losses can be compensated in this fashion. However, this system can hardly be considered as one material and this geometry would not solve the problems mentioned above. It is interesting to note (and relevant for some of the controls in Section 3) that this uncoupled arrangement would always lead to increased transmittance once the gain material is optically pumped. In fact, one would obtain exactly the same differential transmittance change, $\Delta T/T$, with and without the metamaterial, respectively, at any wavelength. For example, if the metamaterial itself transmits 10% of the light and the gain material transmittance changes from 100% to 101% upon optical pumping, the differential transmittance change is $\Delta T/T=1\%$ with and without the lossy metamaterial (*i.e.*, without metamaterial, the overall system transmittance $T$ changes from 100% to 101%; with metamaterial it changes from 10.0% to 10.1%).

This gedankenexperiment makes clear that a coupling between metamaterial layer and gain material is absolutely crucial for success. Coupling in the spirit of a metamaterial can only be achieved if the two structures are not separated by more than the decay length of the evanescent electromagnetic fields – typically 10-20 nm for our conditions.

In the case of weak coupling *via* the evanescent fields, the two resonances do get mixed (or hybridized) to some extent, yet addressing them by their original names remains meaningful. Here, the overall transmittance change is composed of two competing contributions: First, the gain resonance again leads to an increase of transmittance upon optical pumping. Second, the metamaterial resonance (specifically, we consider a spectral transmittance minimum) acquires reduced damping upon optical pumping. Hence, the transmittance minimum narrows up, leading to reduced transmittance at the resonance-frequency position and increased transmittance on the two spectral sides. Which of these two contributions dominates depends on the relative strengths of the two resonances. For the case of a weak gain resonance (that is relevant in the present paper), the second contribution overwhelms the first one and one gets reduced transmittance on resonance. This reasoning, however, is still a bit too naive because a change in gain, hence in the imaginary part of the gain-material refractive index, is necessarily accompanied by a change in the real part of the gain-material refractive index *via* the Kramers-Kronig relations (resulting from causality). Thus, one additionally expects spectral shifts, which are again accompanied by spectral regions of reduced transmittance.

In the case of strong coupling between the two resonances *via* the evanescent fields, the resonances completely loose their original identity. One expects avoided crossings and generally Fano-type resonance lineshapes, including the possibility of complete loss compensation at specific frequencies.

All of the above aspects and limits are included in a simple analytical toy model that has been introduced two years ago [14]. This toy model considers (on a self-consistent footing) two coupled resonances, a Lorentzian resonance of the metamaterial and a second Lorentzian resonance that can be inverted, delivering the gain. The relevance of this toy model to the problem under discussion has explicitly been shown in [14]. We will use this analytical model in Section 4 to fit to our experimental data. It is clear that this toy model leaves lots of space for future improvements regarding theoretical modeling, but it would be very difficult to actually fit a complete numerical model to the vast experimental data to be presented below.

**2. Definition of the Experiment**

The samples in our experiments are fabricated by standard electron-beam lithography on single-crystalline semiconductor wafers that have been grown by molecular-beam epitaxy (MBE) on semi-insulating InP substrates. We choose an operation wavelength at around 1.5 µm because metal losses are already an issue there but losses are not yet as bad as in the visible regime. Our design is the result of investigating several dozens of wafers (single and multiple quantum wells) that we have grown and characterized (and that will not be shown here). For positioning the gain material, one must appreciate that the evanescent fields of split-ring resonators operating at this wavelength and located on such high-refractive-index substrates decay on a scale of 10-20 nm normal to the wafer surface. This number has been obtained from numerical calculations (not shown). This decay length obviously limits the possible thickness of the gain material. The position of the single QW results from a trade-off: On the one hand, the QW should be close to the surface to maximize the coupling to the SRR *via* the evanescent fields. On the other hand, the QW should be sufficiently far away from the surface to prevent evanescent coupling to non-radiative modes ("quenching") or deterioration of the QW optical properties due to substantial overlap of the QW exciton wavefunction with the wafer surface. Furthermore, for the aluminum component involved in our work, the wafer surface needs to be passivated to prevent oxidation. The latter aspects altogether necessitate a minimum separation of the QW from the wafer surface of about 5-10 nm. As expected from our reasoning, we have found that introducing three QWs instead of just one QW does not improve the behavior (not shown).

In regard to obtaining strong coupling between QW and SRR, we note that the optical selection rules for the QW do not work in our favor: The strongest and longest-wavelength interband optical transition, the one involving the heavy-hole valence band, is dipole-forbidden for an electric field oriented perpendicular to the QW plane. Unfortunately, for the SRR used, this field component is expected to be the strongest one within the QW plane. We expect that the combination of these two aspects reduces the effective coupling between SRR and QW.

We have grown the sample shown in Fig. 1(a) in a Riber 32P MBE machine. The InP substrate was degassed at 175° C in vacuum overnight, heated to 490° C under an $As_4$ flux of $1.4 \times 10^{-5}$ Torr, and we have performed the epitaxy at 480-485° C under the same $As_4$ flux. The target layers grown lattice-matched to InP and without any interruptions are: 400-nm $In_{0.52}Al_{0.48}As$ buffer and lower barrier, 12.7-nm $In_{0.53}Ga_{0.47}As$ quantum well, 2.5-nm $In_{0.52}Al_{0.48}As$ upper barrier, and a 2.3-nm $In_{0.53}Ga_{0.47}As$ cap layer. We have measured the surface roughness by atomic-force microscopy to be 0.37 nm root mean square in the center of the 51-mm diameter wafer and 0.54 nm at a radius of 20 mm from the center, indicating smooth interfaces.

Low-temperature photoluminescence spectra of the wafer used throughout this work are shown in Fig. 1(b) for various excitation powers. The photoluminescence spectrum at low excitation exhibits a spectral line full width at half maximum of about 25 nm – a good value for a QW that close to the wafer surface.

Next, we fabricate silver split-ring resonators (SRR) using standard electron-beam lithography and standard high-vacuum electron-beam evaporation of silver onto these wafers. Silver is chosen because of its known low losses compared to other metals. Due to the high-refractive-index semiconductor wafer, the SRR lateral features have to be smaller compared to, *e.g.*, SRR on glass substrates (see, *e.g.*, [2,3]) to reach 1.5-μm operation wavelength. The silver thickness is 30 nm. The individual arrays composed of SRR square lattices with lattice constant $a$=250 nm have a footprint of 100 μm × 100 μm.

On each wafer piece we fabricate an entire set of SRR arrays in which the electron-beam exposure dose is systematically varied. This leads to a variation of the geometrical SRR parameters, hence to a tuning of the SRR resonance wavelength from about 1.3 μm to about 1.8 μm. This lithographic tuning allows us to systematically change the spectral detuning between the QW gain position and the SRR resonance. A typical top-view electron micrograph is shown in Fig. 1(c); the measured room-temperature normal-incidence transmittance spectra of all SRR arrays discussed within this paper are displayed in Fig. 1(d). Normalization of the transmittance is with respect to the bare quantum wells.

For all femtosecond experiments shown in this work, the samples are held at helium temperatures in a microscope helium flow cryostat (KryoVac). The actual sample temperature (without optical excitation) is $T$=5-10 K.

Optical pumping of the QW to achieve gain is accomplished by pumping the samples from the QW side of the wafer with 150-fs optical pulses at around 810-nm center wavelength derived from a Ti:sapphire mode-locked laser oscillator (Tsunami from Spectra-Physics, 81-MHz repetition frequency). This oscillator also pumps an optical parametric oscillator OPO (Opal from Spectra-Physics) that is tunable in the spectral range from 1.4 μm to 1.6 μm wavelength. The samples have been designed to be centered in this OPO interval. Pump and probe pulses with variable time delay, $\Delta t$, are coaxially focused onto the samples by means of a single 5-cm focal length lens. To ensure that the pump spot is sufficiently large compared to the probe spot to obtain spatially homogeneous excitation conditions, we use two independently adjustable optical telescopes for the pump and the probe beam, respectively. In the sample plane, using a knife-edge technique, we measure a pump-spot diameter of 22 μm (full width at half maximum) and a probe-spot diameter of 10 μm. These spots are sufficiently smaller than the footprint of the SRR arrays quoted above to avoid edge effects. To monitor the focusing conditions as well as the alignment of the two spots relative to each other and relative to the SRR arrays, we image the sample plane onto a PbS camera.

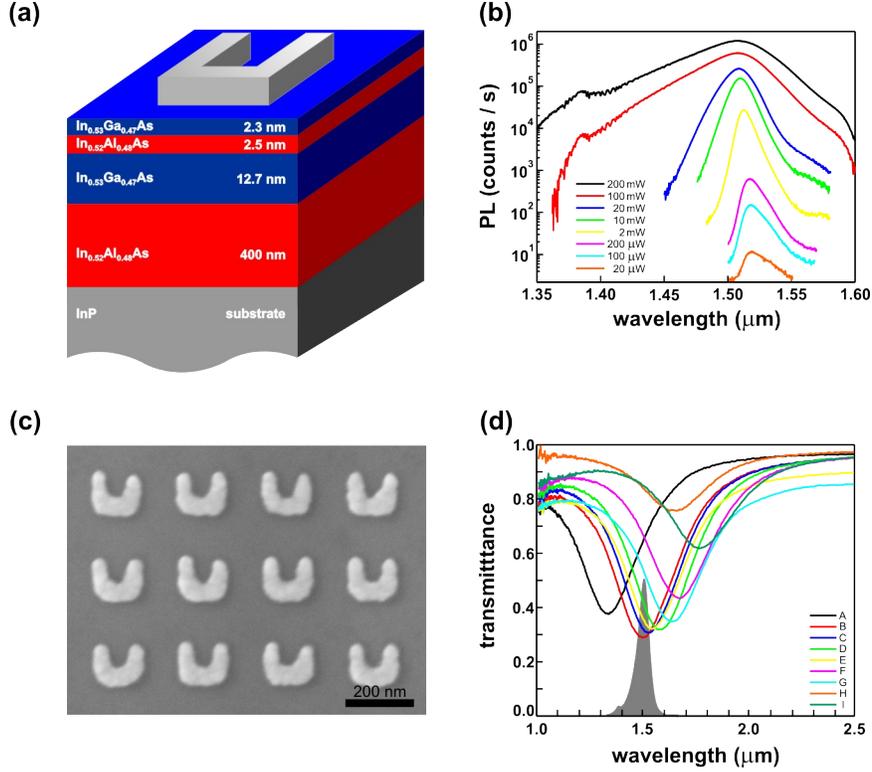

Fig. 1. (a) Layer sequence (not to scale) and composition of the semiconductor crystal structure grown by molecular-beam epitaxy (MBE), lattice-matched to the InP substrate. After MBE growth, arrays of 30-nm thin Ag split-ring resonators (SRR) are fabricated on top of this wafer using a standard electron-beam-lithography process. (b) Quantum well photoluminescence (PL) spectra for increasing power, $P$, of the exciting femtosecond Ti:sapphire laser at 0.81-µm center wavelength as indicated. In the pump-probe experiments depicted in Figs. 2-4, we have used the identical laser with $P$=200 mW under identical focusing conditions. (c) Electron micrograph of SRR array "D". (d) Optical normal-incidence linear intensity transmittance spectra of some of the SRR arrays used in this work (horizontal linear incident polarization of light). The gray area corresponds to the PL spectrum of the bare QW for an excitation power of $P$=200 mW.

The transmitted probe beam is collected, spectrally filtered to suppress the residual pump, and sent onto a room-temperature Ge photodetector. The pump beam is chopped at about 0.4-kHz frequency and the differential signal $\Delta T$ is detected using a standard Lock-In technique. By blocking the probe beam, we ensure that only a negligible fraction of that differential signal stems from pump-induced photoluminescence (that cannot really be spectrally filtered). Upon additionally measuring the probe transmittance, $T$, without pump, we obtain the relative transmittance change $\Delta T/T$. We follow the usual sign convention, *i.e.*, $\Delta T/T>0$ corresponds to increased sample transmittance upon optical pumping. The pump polarization is horizontal with respect to the SRR shown in Fig. 1(c), the linear probe polarization is always varied between horizontal and vertical polarization to provide controls. The meaning of these controls will be discussed in Section 3.

We also perform all experiments on the SRR fields and soon thereafter on the side on a region with only the bare QW to provide another set of controls. The meaning of these controls will also be discussed in Section 3. Such experiments are performed for many OPO wavelengths.

As further control experiments, we have twice removed the single QW by mechanical polishing and have performed similar pump-probe experiments on the remaining InP wafers. We have found no detectable differential transmittance signals at all, indicating that no nonlinear contribution outside the noise level stems from the InP substrate for all data shown in this work. Thus, the thick InP substrate can be considered as a passive linear dielectric for the purpose of the present paper.

The average power of the probe beam is set to not more than 0.07 mW in front of the sample (clearly the OPO power varies with OPO wavelength). By further attenuating the probe power at the price of worse signal-to-noise ratio, we have verified that the probe is within the linear regime. The pump beam power is fixed at about 200 mW in front of the sample in this work – unless stated otherwise. This is the maximum power accessible under our experimental conditions (the main part of the Ti:sapphire laser oscillator pumps the OPO). Notably, we have not observed any sample deterioration whatsoever under these conditions – not even over many days. This pump power is expected to completely saturate the QW. We find that the QW photoluminescence spectrum develops a pronounced high-energy tail at powers much below 200 mW [see Fig. 1(b)]. Below, we will discuss on selected examples that the general behavior does not depend on the pump power, whereas the absolute signal strengths obviously do. This observation clearly implies that we have not found any indications of lasing (or "spasing") in our work.

## 3. Low-Temperature Femtosecond Pump-Probe Experiments

Figure 2 gives an overview over the data obtained from one of the SRR arrays, namely one that is nearly resonant with the expected peak of the QW gain spectrum (compare Fig. 1). Data for horizontal (left column in Fig. 2) and vertical (right column in Fig. 2) OPO probe polarization are depicted. For horizontal polarization, the light does couple to the fundamental SRR resonance, for vertical polarization it does not. In all cases, the red curves correspond to measurements on the SRR arrays, the blue curves to measurements on the bare QW on the side, *i.e.*, without any SRR array. The OPO probe center wavelength increases from top to bottom. Note that the width of the OPO spectrum is about 20 nm. Hence, the data have been taken with 20-nm spectral separation. In each case, the dashed black horizontal line defines $\Delta T/T$=0.

Inspection of the left column of Fig. 2 shows a rather different behavior for the SRR on the QW compared to the bare QW. While the bare QW always delivers positive $\Delta T/T$ signals below about +2% (blue curves), the sign and magnitude of the signals change for the SRR on the QW (red curves). Under some conditions, $\Delta T/T$ reaches values as negative as -8%. Following our introduction, this behavior can be expected for the coupled system of SRR and QW.

In addition to the sign and magnitude of the differential transmittance signal, the temporal behavior is also quite different for the case of SRR on the QW compared to the QW alone. Precisely, the temporal decays tend to become much faster with SRR compared to the case without. For example, for an OPO probe wavelength of 1.48 µm, the blue measured curve can nicely be fitted with a single-exponential decay with a time constant of 670 ps (see dots in Fig. 2). The red curve cannot be fitted with a single exponential (not shown). It can, however, be fitted with the sum of two exponentials (see dots in Fig. 2), one with time constant 15 ps and the other with time constant 180 ps.

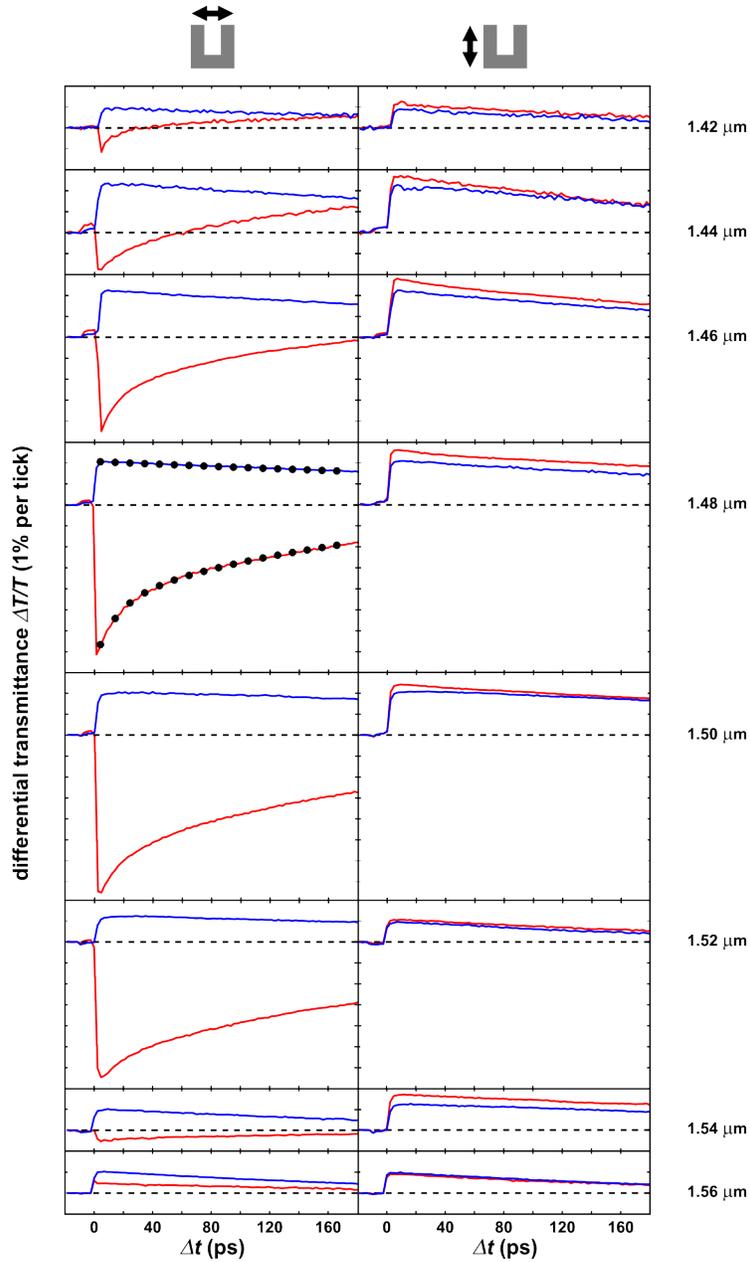

Fig. 2. Femtosecond pump-probe experiments for an array of SRR that is resonant with the QW gain (sample "D" in Figs. 1 and 4). The left column corresponds to horizontal probe polarization with respect to the SRR, the right column to vertical probe polarization. The differential transmittance signals for the case of QW and SRR are shown by the red curves, the case of QW alone by the blue curves. Zero differential signal, *i.e.*, $\Delta T/T=0$, is indicated by the dashed horizontal lines. One tick on the vertical axis corresponds to $\Delta T/T=1\%$. The curves are unequally vertically displaced for clarity. The OPO probe wavelength increases from top to bottom as indicated on the right-hand side. Sample temperature is 5-10 K, the 150-fs pump pulses are centered at around 0.81-µm wavelength

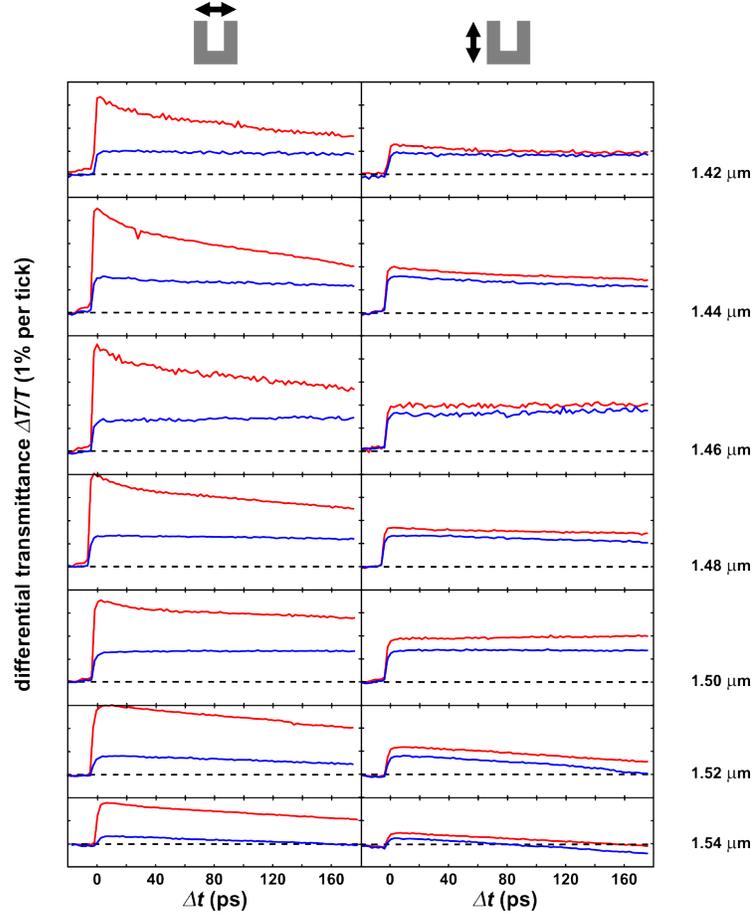

Fig. 3. Same as Fig. 2, but for an off-resonant array of split-ring resonators (sample "A" in Figs. 1 and 4).

This fast decay time contribution might have three different origins: (i) quenching due to the nearby metal of the SRR, (ii) emergence of lasing (or spasing), or (iii) the Purcell effect. Possibility (i) can be tested on the basis of the control experiments shown on the right-hand column of Fig. 2. There, the light does not couple to the fundamental SRR resonance with vertical polarization. Quenching, *i.e.*, coupling to non-radiative SRR modes would clearly persist. The experiment rather shows hardly any difference between the blue and red curves in the right column – unambiguously ruling out explanation (i). Hypothesis (ii) would imply a significant dependence of the temporal decay on the pump power, *i.e.*, on the QW inversion. In additional experiments (not shown), we have varied the pump power (while keeping the ratio of pump to probe power fixed). This obviously leads to smaller differential transmittance signals, but the temporal behavior with the two time constants remains unchanged. This finding definitely rules out explanation (ii). This leaves us with possibility (iii), *i.e.*, the Purcell effect. Indeed, a local intensity enhancement of a factor of 10-50 is not unusual at all in the field of plasmonics. Intuitively, the rapid-temporal-decay contribution could originate from those regions of the QW directly under and/or nearby the Ag SRR, the slow-temporal-decay contributions from the QW regions in between the SRR (see electron micrograph in Fig. 1). Notably, off resonance, the temporal decays of QW with SRR and QW without SRR become more similar in Fig. 2. Indeed, for these off-resonant conditions, one expects less local intensity enhancement, hence a less pronounced Purcell effect.

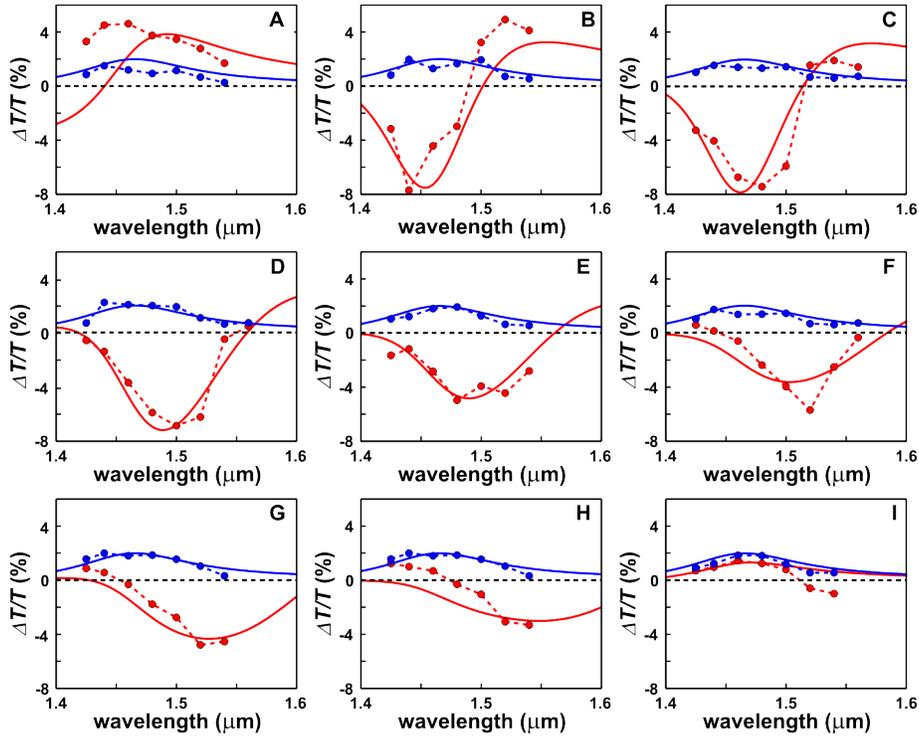

Fig. 4. Summary of femtosecond pump-probe experiments (like exemplified in Figs. 2 and 3) for various SRR arrays. The OPO probe polarization is horizontal. The array letters given in the upper right-hand corners correspond to those in Fig. 1(d). The color coding is the same as in Figs. 2 and 3. Spectra of the differential transmittance signal are shown versus OPO probe wavelength for a fixed time delay of $\Delta t=+5$ ps. Dots connected by dashed straight lines correspond to the experiment, the solid curves are derived from the toy model (parameters are given in the main text). For reference, the measured linear optical spectra of the corresponding SRR arrays are shown in Fig. 1(d).

The behavior of the magnitude of the differential transmittance signal in Fig. 2 is also consistent with this interpretation. For a QW with SRR, the modulus of the signals is a factor of about 5 larger than for the case of the bare QW. In Section 4, we will show that this measured behavior is consistent with the toy model, the physics of which has been explained in the introduction (Section 1).

We have taken complete data sets similar to that in Fig. 2 for several other SRR arrays. A second example is shown in Fig. 3. Here, the SRR resonance is largely blue-detuned with respect to the QW gain spectrum, resulting in an off-resonant situation. To give an overview on all these results, we compress the data as shown in Fig. 4. Here, we use the same color coding as in Figs. 2 and 3 and merely plot the differential transmittance signal at small positive time delay, *i.e.*, for $\Delta t=+5$ ps, versus the OPO probe wavelength. It becomes obvious that negative $\Delta T/T$ signals for the case of QW and SRR are the rule rather than the exception. In contrast, the $\Delta T/T$ signals are always positive (and generally smaller in magnitude) for the bare QW. Close to resonance between the SRR resonance and the QW gain peak, the behavior is similar to that shown in Fig. 2. For more off-resonant SRR arrays, the differences between QW and SRR and QW alone become smaller – as expected from resonant coupling between the SRR resonance and the QW gain.

## 4. Comparison with Toy Model

To obtain a better understanding for the fairly complex measured behavior shown in Figs. 2-4, we compare these data with the above-mentioned toy model that has been defined in detail in [14]. We refer the reader to the mathematics therein. In Section 1 of the present paper, we have already revisited the underlying physics.

Clearly, this toy model [14] has quite a few free parameters: For the resonances of SRR and QW, respectively, one can choose the resonance position $\Omega$, the damping $\gamma$, the number density $N$ and the dipole matrix element $d$ – four parameters each. In addition, the local-field coupling parameter $L$ is purely heuristic. Altogether, this makes 9 adjustable parameters. However, the four parameters for the QW gain resonance can already be adjusted on the basis of the experiments on the bare QW (see Fig. 4) where the background dielectric constant is set to $(3.1)^2$. The number density of the SRR can be taken from Fig. 1(c), the other three parameters for the SRR resonance without gain (*i.e.*, without optical pumping) can already be adjusted to fit the linear optical spectra of the SRR arrays shown in Fig. 1(d). The combination of these boundary conditions essentially leaves the local-field parameter $L$ as the only truly free parameter. It is expected to be similar for all SRR arrays, regardless of their resonance wavelengths.

In more detail, we compute the differential transmittance by taking the difference between the transmittance of the pumped case (assuming $f=1$ occupation of the upper level) and the unpumped case (*i.e.*, $f=0$) and normalize the result with respect to the unpumped transmittance. Using the Maxwell-Garnett approximation described in [14], the 30-nm thin layer of SRR and the 12.7-nm thin single QW are treated as a single effective layer with a thickness of 42.7 nm on top of a dielectric halfspace, the InP substrate. The InP refractive index are taken as $n=3.1$ (independent of wavelength). In principle, the spectra obtained along these lines should finally be convoluted with the about 20-nm broad OPO probe spectrum. However, for the present conditions, the resulting effects turn out to be unimportant (not shown). Thus, we avoid this process step here. For all samples "A" to "I", the following model parameters (same nomenclature as in [14]) are fixed: $\Omega_{2LS}=2\pi\times205$ THz, $\gamma_{2LS}=50$ THz, $d_{2LS}=6.5\times10^{-29}$ Cm, $N_{2LS}=2.1\times10^{24}$ m$^{-3}$, $\gamma_{pl}=90$ THz, $d_{pl}=4.2\times10^{-26}$ Cm, and $N_{pl}=5.3\times10^{20}$ m$^{-3}$. The plasmonic resonance frequencies $\Omega_{pl}$ and the local-field parameters $L$ for all samples shown in this paper are listed below.

Table 1. Model parameters $\Omega_{pl}$ and $L$ that are individually adjusted for samples "A" to "I" (compare Fig. 4).

| sample | $\Omega_{pl}$ | $L$ |
|---|---|---|
| A | $2\pi\times241$ THz | $1.8\times10^{10}$ F/m |
| B | $2\pi\times205$ THz | $1.7\times10^{10}$ F/m |
| C | $2\pi\times203$ THz | $1.8\times10^{10}$ F/m |
| D | $2\pi\times195$ THz | $2.1\times10^{10}$ F/m |
| E | $2\pi\times193$ THz | $1.8\times10^{10}$ F/m |
| F | $2\pi\times190$ THz | $2.1\times10^{10}$ F/m |
| G | $2\pi\times187$ THz | $2.2\times10^{10}$ F/m |
| H | $2\pi\times183$ THz | $2.1\times10^{10}$ F/m |
| I | $2\pi\times175$ THz | $0.1\times10^{10}$ F/m |

Corresponding results are also shown in Fig. 4, allowing for direct comparison with experiment. Obviously, the general agreement between experiment and toy model is excellent considering the complexity of the overall behavior. In particular, we again find that negative differential transmittance signals $\Delta T/T$ are the rule rather than the exception. These negative signals are mainly due to reduction of the damping of the SRR transmittance minimum (see discussion in Section 1). Furthermore, in agreement with experiment, we find that the magnitude of the $\Delta T/T$ signals is substantially larger for the case of QW and SRR compared to the case of the QW alone. Let us emphasize, as already discussed in Section 1, that without any coupling between SRR and QW, *i.e.*, for local-field parameter $L=0$, the $\Delta T/T$ signals for the case of QW and SRR and the case of QW alone would be just identical (not shown) – in striking disagreement with the experimental facts. This once again emphasizes that our results imply a considerably strong local-field coupling between the SRR and the QW gain – which is at the heart of our aim of reducing the metamaterial's losses.

Let us note in passing that our simple toy model is completely unable to properly describe the Purcell effect. Precisely, if we let the upper-state two-level-system occupation $f$ relax exponentially from $f=1$ to $f=0$, all $\Delta T/T$ signals with and without SRR show the same exponential decay. This is due to the fact that the Purcell effect is a quantum optical phenomenon, whereas our toy model is semi-classical.

Finally, it is interesting to ask on the basis of the toy model: How far are we away from the original aim of completely compensating metamaterial losses? Let us consider the best case, *i.e.*, that the two-level system (2LS) resonance frequency $\Omega_{2LS}$ representing the QW and the plasmonic resonance (pl) frequency $\Omega_{pl}$ representing the SRR array are degenerate. Under this condition, a simple and transparent way to summarize the influence of the various parameters is to start from the lasing spasing condition (Eq. (13) in [14]) for the two-level system upper-state occupation $f$, *i.e.*, from

$$f = \frac{1}{2}\left(1 + \frac{\gamma_{pl}\gamma_{2LS}}{V_{pl}V_{2LS}}\right) =: \frac{1}{2}\left(1 + \frac{\gamma^2}{V^2}\right) \in [0,1]. \quad (1)$$

Here, $\gamma_{2LS}$ and $\gamma_{pl}$ are the damping frequencies of the two-level system and the plasmonic resonance. The quantities $V_{2LS}$ and $V_{pl}$ are the respective effective coupling frequencies given by (Eq. (9) in [14])

$$\begin{aligned}V_{2LS} &= \hbar^{-1}d_{2LS}LN_{pl}d_{pl}, \\ V_{pl} &= \hbar^{-1}d_{pl}LN_{2LS}d_{2LS}.\end{aligned} \quad (2)$$

The condition (1) is graphically illustrated in Fig. 5. In the gray triangle, lasing spasing is *not* possible because $f$ has a mathematical solution outside the interval [0,1]. In other words, the available gain is simply not sufficient to compensate for the losses. In the white triangle, lasing spasing is possible, even under continuous-wave self-consistent conditions. The red dot corresponds to the parameters that we have used to fit to the experimental data in Fig. 2.

To move the red dot into the white triangle, one could, *e.g.*, increase the QW dipole moment $d_{2LS}$ by a factor of three to four ("more gain"). Alternatively, one could increase the local field coupling parameter $L$ by the same factor ("more effective metamaterials at fixed QW gain"), or reduce the SRR damping ("less loss"), or reduce the QW damping ("more gain") by the same factor. On the basis of the experimental results of this paper, these parameter improvements are not quite impossible, but are not simple to reach either. Other material systems may have to be chosen.

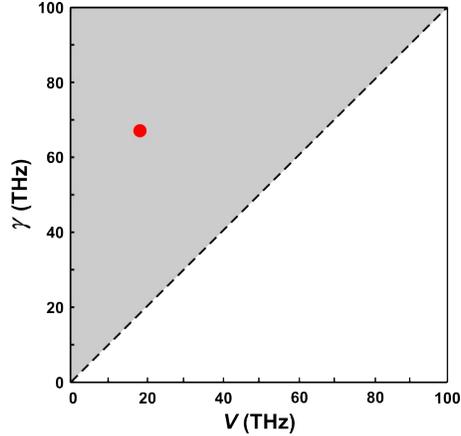

Fig. 5. Illustration of the lasing (spasing) condition for the case that the quantum-well and the split-ring-resonator resonances are degenerate [see Eq. (1)]. Lasing and complete loss compensation are possible within the white triangle, but are not possible in the gray triangle. The two triangles are separated by the straight line with $\gamma=V$ [see Eq. (1)]. The red dot inside the gray triangle corresponds to the experimental conditions of sample "D" and is obtained from exactly the same model parameters that we have used to fit to these experimental results (compare Fig. 4).

## 5. Conclusions

We have presented experiments aiming at compensating the metal losses of arrays of SRR by coupling to an optically pumped InGaAs single quantum well *via* the local (or evanescent) electromagnetic fields of the SRR. We observe differential transmittance signals from the coupled system that are a factor of four to five larger than for the bare quantum well. Furthermore, we observe a more rapid temporal decay of the differential transmittance signal for the coupled system compared to the bare quantum well (Purcell effect). Both effects indicate substantial local-field-enhancement effects, which increase the effective metamaterial gain beyond the bare quantum-well gain, leading to a significant reduction of the metamaterial's losses. This interpretation is also confirmed by comparison of the experimental data with a recently introduced analytical toy model. However, despite of the fact that we have employed very intense pulsed optical pumping and that we have cooled the samples to helium temperatures and that we have optimized the semiconductor-wafer design, the magnitude of the effect is too small to obtain complete metamaterial-loss compensation in our experiments.

**Acknowledgements**

We acknowledge support by the Deutsche Forschungsgemeinschaft (DFG) and the State of Baden-Württemberg through the DFG-Center for Functional Nanostructures (CFN) within subproject A1.5. The project PHOME acknowledges the financial support of the Future and Emerging Technologies (FET) programme within the Seventh Framework Programme for Research of the European Commission, under FET-Open grant number 213390. The project METAMAT is supported by the Bundesministerium für Bildung und Forschung (BMBF). The PhD education of N.M and M.R. is embedded in the Karlsruhe School of Optics & Photonics (KSOP). Work at Ames Lab was supported by Dept. of Energy (Basic Energy Sciences), contract No. DE-AC02-07CH11358. The Tucson group thanks AFOSR and NSF AMOP for support. HMG and JH thank the Alexander von Humboldt Foundation for a Renewed Research Stay and Junior Scientist Award, respectively.